\documentclass[conference]{IEEEtran}

\usepackage{epsfig,amsmath,amssymb,latexsym}
\usepackage{graphicx,epsfig,color,epsf,psfrag,hhline,amsmath,amssymb,textcomp}

\usepackage[ruled]{algorithm2e}

\def\_{\rule{.3em}{.15ex}}      % Get underscore by typing \_.

 \normalmarginpar
\newcommand {\mymarginpar}[1]{\marginpar{#1}}
\renewcommand {\marginpar}[1]{}

\def\_{\rule{.3em}{.15ex}}      % Get underscore by typing \_.

\newcommand{\ls}[1]
   {\dimen0=\fontdimen6\the\font
    \lineskip=#1\dimen0
    \advance\lineskip.5\fontdimen5\the\font
    \advance\lineskip-\dimen0
    \lineskiplimit=.9\lineskip
    \baselineskip=\lineskip
    \advance\baselineskip\dimen0
    \normallineskip\lineskip
    \normallineskiplimit\lineskiplimit
    \normalbaselineskip\baselineskip
    \ignorespaces
   }
\newcommand {\bearn}{\begin{eqnarray*}}
\newcommand {\eearn}{\end{eqnarray*}}
\newcommand {\barr}{\begin{array}}
\newcommand {\earr}{\end{array}}

\newcommand {\N}{{\cal N}}

\newtheorem{definition}{Definition}
\newtheorem{property}[definition]{Property}
\newtheorem{proposition}[definition]{Proposition}
\newtheorem{lemma}[definition]{Lemma}
\newtheorem{theorem}[definition]{Theorem}
\newtheorem{corollary}[definition]{Corollary}
\newtheorem{example}[definition]{Example}
\newtheorem{remark}[definition]{Remark}
\newcommand {\benum} {\begin{enumerate}}
\newcommand {\eenum} {\end{enumerate}}

\newcommand {\bdesc} {\begin{description}}
\newcommand {\edesc} {\end{description}}

\newcommand {\bfig}[2] {\begin{figure}
  \centering
  \includegraphics[width=#2]{#1}}
\newcommand {\brotatefig}[2] {\begin{figure}[htbp]
                        \centerline {
                         \epsfig{figure={#1},clip=,angle=-90,width={#2}}}}
\newcommand {\bfigfirst}[2] {\begin{figure}[h]
                        \centerline {
                        \setlength{\epsfxsize}{#2}
                        \epsffile{#1}}}
\newcommand {\efig}[2]{ \caption{#2}
                        \label{fig:#1}
                        \end{figure}
                        \mymarginpar{fig:#1}}
\newcommand {\erotatefig}[2]{ \caption{#2}
                        \label{fig:#1}
                        \end{figure}
                        \mymarginpar{fig:#1}}
\newcommand {\rfig}[1]{Figure \ref{fig:#1}}

\newcommand {\btab}[1]{
                       \begin{table}
                       \centering
                       \begin{tabular}{#1}}
\newcommand {\etab}[3] {
                       \end{tabular}
                       \caption[#3]{#2}
                       \label{tab:#1}
                       \end{table}
                       \mymarginpar{tab:#1}
                       \vspace{.1in}}

\newcommand {\btabular}[1]{\begin{center}
                       \begin{tabular}{#1}}
\newcommand {\etabular}{\end{tabular}
                       \end{center}}

\newcommand {\bdefin}[1]{\begin{definition}
                      \mymarginpar{def:#1}
                      \label{def:#1} }
\newcommand {\edefin}       {\end{definition}}
\newcommand {\rdef}[1]{Definition \ref{def:#1}}

\newcommand {\bpro}[1]{\begin{property}
                      \mymarginpar{pro:#1}
                      \label{pro:#1} }
\newcommand {\epro}   {\end{property}}

\newcommand {\bprop}[1]{\begin{proposition}
                      \mymarginpar{prop:#1}
                      \label{prop:#1} }
\newcommand {\eprop}       {\end{proposition}}
\newcommand {\rprop}[1]{Proposition \ref{prop:#1}}

\newcommand {\blem}[1]{\begin{lemma}
                      \mymarginpar{lem:#1}
                      \label{lem:#1} }
\newcommand {\elem}   {\end{lemma}}
\newcommand {\rlem}[1]{Lemma \ref{lem:#1}}

\newcommand {\bthe}[1]{\begin{theorem}
                      \mymarginpar{the:#1}
                      \label{the:#1} }
\newcommand {\ethe}   {\end{theorem}}
\newcommand {\rthe}[1]{Theorem \ref{the:#1}}

\newcommand {\bcor}[1]{\begin{corollary}
                      \mymarginpar{cor:#1}
                      \label{cor:#1} }
\newcommand {\ecor}   {\end{corollary}}

\newcommand {\bax}[1]{\begin{axiom}
                      \mymarginpar{ax:#1}
                      \label{ax:#1} }
\newcommand {\eax}       {\vspace{-.1in} \end{axiom}}

\newcommand {\bex}[2]{\vspace{.1in}
                      \begin{example}
                      \mymarginpar{ex:#1}
                       {\bf #2}
                      \label{ex:#1} }
\newcommand {\eex}       {\end{example} \vspace{.3cm} }

\newcommand {\brem}[1]{\begin{remark}
                      \mymarginpar{rem:#1}
                      \label{rem:#1} \em }
\newcommand {\erem}   {\end{remark}}

\newcommand {\beq}[1]{\mymarginpar{eq:#1}
                      \begin{equation}
                      \label{eq:#1} }

\newcommand {\beqno}[1]{\mymarginpar{eq:#1}
                      \begin{eqnarray}
                      \nonumber}

\newcommand {\eeq}       {\end{equation}}
\newcommand {\eeqno}       { && \end{eqnarray}}
\newcommand {\req}[1]{(\ref{eq:#1})}

\newcommand {\bear}[1]{\mymarginpar{eq:#1}
                       \begin{eqnarray}
                       \label{eq:#1} }

\newcommand {\bearno}[1]{\mymarginpar{eq:#1}
                       \begin{eqnarray}
                       \nonumber}

\newcommand {\eear}{\end{eqnarray}}
\newcommand {\eearno}{\end{eqnarray}}
\newcommand {\bsel}{\left \{ \begin{array}{cl}}
\newcommand {\esel}{\end{array} \right.}

\newcommand {\bmat}[1]{\left [ \begin{array}{#1}}
\newcommand {\emat}{\end{array} \right ]}
\newcommand {\bsec}[2]{\mymarginpar{sec:#2}
                       \section{#1}
                       \label{sec:#2} }

\newcommand {\bsubsec}[2]{\mymarginpar{sec:#2}
                       \subsection{#1}
                       \label{sec:#2} }

\def\R{I\kern-0.30em R}
\def\N{I\kern-0.30em N}
\def\P{I\kern-0.30em P}

\newcommand\gbar{{\bar g}}
\newcommand\dbar{{\bar d}}
\newcommand\gsim{g}
\newcommand\tgsim{{\tilde g}}

\newcommand\Deltaw{\Delta_{\mbox{a}}}
\newcommand\tDeltaw{\tilde \Delta_{\mbox{a}}}
\newcommand\ksetsplus{$\mbox{K-sets}^+$ }
\newcommand\ksets{K-sets }
\newcommand\seta{1}
\newcommand\setb{2}

\begin{document}

%\baselineskip24.8pt

%\title{$\mbox{K-sets}^+$: an Iterative Algorithm for Clustering in a Semi-metric Space}
\title{$\mbox{K-sets}^+$: a Linear-time Clustering Algorithm for Data Points with a Sparse Similarity Measure}
%\title{\ksetsplus: An Iterative  Algorithm Beyond $K$-means}
%Clustering in Non-Euclidean Spaces}
%\title{A Markov Chain Approach for Relative Centrality and Community Detection}

%\title{From Newman's Fast Algorithm to a General Probabilistic Framework for Detecting Community Structure in %Networks}

\author{ Cheng-Shang Chang,~Chia-Tai Chang,~Duan-Shin Lee  and  Li-Heng Liou\\
Institute of Communications Engineering\\
National Tsing Hua University \\
Hsinchu 30013, Taiwan, R.O.C. \\
Email:  cschang@ee.nthu.edu.tw; s104064540@m104.nthu.edu.tw; lds@cs.nthu.edu.tw; dacapo1142@gmail.com  \\
%g9764514@oz.nthu.edu.tw; tm629199@yahoo.com.tw
% lds@cs.nthu.edu.tw
% \thanks{This paper was presented in part at the IEEE International Conference
%        on Computer Communications (INFOCOM'11), Shanghai, China, 2011.}
%\thanks{This research was supported in part by the National Science Council,
%Taiwan, R.O.C., under Contract NSC-94-2213-E-007-046, Contract
%NSC-95-2221-E-007-039, and the Program for Promoting Academic
%Excellence of Universities NSC 94-2752-E-007-002-PAE.}
}

%\date{(July 13, 2007)}

%\markboth{Draft: May 11, 2001}
%{Murray and Balemi: Using the style file IEEEtran.sty} %!PN
%{Murray and Balemi: Using the Document Class IEEEtran.cls} %!PN

\maketitle

\begin{abstract}
%\baselineskip18pt %
%Based on Newman's fast algorithm \cite{Newman04},

In this paper, we first propose a new iterative algorithm, called the \ksetsplus algorithm for clustering data points in a semi-metric space, where the distance measure does not necessarily satisfy the triangular inequality. We show that the \ksetsplus algorithm converges in a finite number of iterations and it retains the same performance guarantee as the K-sets algorithm for clustering data points in a metric space. We then extend the applicability of the \ksetsplus algorithm from data points in a semi-metric space to data points that only have a symmetric similarity measure. Such an extension leads to great reduction of computational complexity. In particular, for an $n \times n$ similarity matrix with $m$ nonzero elements in the matrix, the computational complexity of the \ksetsplus algorithm is $O((Kn+m)I)$, where $I$ is the number of iterations. The memory complexity to achieve that computational complexity is $O(Kn+m)$.
As such,  both the computational complexity and the memory complexity are linear in $n$ when the $n \times n$ similarity matrix is {\em sparse}, i.e., $m=O(n)$. We also conduct various experiments to show the effectiveness of the \ksetsplus algorithm by using a synthetic dataset from the stochastic block model and a real network from the WonderNetwork website.
\end{abstract}

%\begin{keywords}

{\bf keywords:} Clustering; community detection
%\end{keywords}
%\fi
%\vfill\eject \pagestyle{plain} \pagestyle{headings}
%\markboth{Draft: June 27, 2003}{}
%\setcounter{page}{100}
%\pagenumbering{arabic}
%\baselineskip 18pt

%\pagestyle{plain} \pagestyle{headings}

\bsec{Introduction}{introduction}

The problem of clustering is of fundamental importance to data analysis  and it has been  studied extensively in the  literature (see e.g., the books \cite{theodoridispattern,rajaraman2012mining}
and the historical review papers \cite{jain1999data,jain2010data}). In such a problem, there is a set of data points  and a similarity
 (or dissimilarity) measure that measures how similar two data points are. The objective of a clustering algorithm is to cluster the data points so that data points within the same cluster are similar to each other and data points in different clusters are dissimilar.
 Clustering is in general considered as an ill-posed problem and there are already many clustering algorithms proposed in the literature, including  the hierarchical algorithm \cite{theodoridispattern,rajaraman2012mining}, the K-means algorithm \cite{jain2010data,lloyd1982least,agarwal2013k}, the K-medoids algorithm \cite{theodoridispattern,kaufman2009finding,van2003new,park2009simple}, the  kernel and spectral clustering algorithms  \cite{shi2000normalized,von2007tutorial,filippone2008survey,krzakala2013spectral}, and the definition-based algorithms \cite{ester1996density,cuevas2001cluster,halkidi2008density,balcan2013clustering}. However, clustering theories that justify the use of these clustering algorithms are still unsatisfactory.

Recently, a mathematical clustering theory was developed in \cite{chang2015mathematical} for clustering data points in a metric space. In that theory, clusters can be formally defined and stated in various equivalent forms. In addition to the definition of a cluster in a metric space,  the \ksets algorithm was proposed in \cite{chang2015mathematical} to cluster data points in a metric space.
The key innovation of the \ksets algorithm in \cite{chang2015mathematical} is the triangular distance that measures the distance from a data point to a set (of data points) by using the triangular inequality.
Like the K-means algorithm, the \ksets algorithm is an iterative algorithm that repeatedly assigns every data point to the closest set in terms of the triangular distance. It was shown in \cite{chang2015mathematical} that the \ksets algorithm converges in a finite number of iterations and outputs $K$ disjoint sets such that any two sets of these $K$ sets are two disjoint clusters when they are viewed in isolation.

 The first contribution of this paper is to extend the clustering theory/algorithm  in
\cite{chang2015mathematical} to data points in a semi-metric space, where the distance measure does not necessarily satisfy the triangular inequality. Without the triangular inequality,
 the triangular distance in the \ksets algorithm is no longer nonnegative and thus the \ksets algorithm may not converge at all.
 Even if it converges, there is no guarantee that the output of the \ksets algorithm are clusters.
To tackle this technical challenge,
we propose
 the \ksetsplus algorithm
 for clustering in a semi-metric space. In the \ksetsplus algorithm, we need to modify the original definition of the triangular distance
 so that the nonnegativity requirement of the triangular distance can be lifted.  For this, we propose the
  adjusted  triangular distance (in \rdef{kdn}) and show (in \rthe{main}) that the \ksetsplus algorithm  that repeatedly assigns every data point to the closest set in terms of the adjusted triangular distance converges in a finite number of iterations. Moreover, the \ksetsplus algorithm  outputs $K$ disjoint sets such that any two sets of these $K$ sets are two disjoint clusters when they are viewed in isolation.

The second contribution of this paper is to further extend the applicability of the \ksetsplus algorithm from data points in a semi-metric space to data points that only have a symmetric similarity measure. A similarity measure is generally referred to as a bivariate function that measures how similar two data points are.
We show there is  a natural mapping from a symmetric similarity measure to a distance measure in a semi-metric space and the
the \ksetsplus algorithm that uses this  distance measure converges to the same partition as that using the original symmetric similarity measure. Such an extension leads to great reduction of computational complexity for the \ksetsplus algorithm.
%as a similarity measure can be very sparse. 
For an $n \times n$ similarity matrix with only $m$ nonzero elements in the matrix,
we show that
 the computational complexity of the  \ksetsplus algorithm  is  $O((Kn+m)I)$, where $I$ is the number of iterations. The memory complexity to achieve that computational complexity is $O(Kn+m)$. If the $n \times n$ similarity matrix is {\em sparse}, i.e., $m=O(n)$, then both the computational complexity and the memory complexity are linear in $n$.

To evaluate the performance of the \ksetsplus algorithm, we conduct two experiments: (i) community detection of signed networks generated by the stochastic block model, and (ii)
 clustering of a real network   from the WonderNetwork website \cite{Wonder}. Our experiments show that the \ksetsplus algorithm is very effective in recovering the ground-truth edge signs even when the signs of a certain percentage of edges are flipped. For the real network of servers, the \ksetsplus algorithm yields various interesting observations from the clustering results obtained by using the geographic distance matrix  and the latency matrix.

\bsec{Clustering in a semi-metric space}{semimetric}

In this paper, we consider the clustering problem for data points in a semi-metric space.
 Specifically, we
 consider a set of $n$ data points, $\Omega=\{x_1, x_2, \ldots, x_n\}$ and a distance measure $d(x,y)$ for any two points $x$ and $y$ in $\Omega$.
The distance measure $d(\cdot, \cdot)$  is assumed to be a {\em semi-metric} and it satisfies the following three properties:
\begin{description}
\item[(D1)] (Nonnegativity) $d(x,y) \ge 0$.
\item[(D2)] (Null condition) $d(x,x)=0$.
\item[(D3)] (Symmetry) $d(x,y)=d(y,x)$.
\end{description}

The semi-metric assumption is weaker than the metric assumption in \cite{chang2015mathematical}, where the distance measure is assumed to satisfy the triangular inequality. In \cite{chang2015mathematical}, the \ksets algorithm was proposed for clustering data points in a metric space.
One of the main contributions of this paper is to propose the \ksetsplus algorithm (as a generalization of the \ksets algorithm) for clustering data points in a {\em semi-metric} space. As  both Euclidean spaces and metric spaces are spacial cases of semi-metric spaces, such a generalization allows us to unify the
well-known K-means algorithm and  the \ksets algorithm in \cite{chang2015mathematical}.

\bsubsec{Semi-cohesion measure}{semi}

Given a semi-metric
$d(\cdot,\cdot)$  for $\Omega$, we define the induced semi-cohesion measure as follows:
\bear{csim7777}
\gsim(x,y)&=&{1 \over n}\sum_{z_2 \in \Omega} d(z_2,y)+{1 \over n} \sum_{z_1 \in \Omega}d(x,z_1)\nonumber\\
&-&{1 \over n^2} \sum_{z_2 \in \Omega}\sum_{z_1 \in \Omega}d(z_2,z_1)-d(x,y).
\eear
It is easy to verify that the induced semi-cohesion measure
satisfies
the following three properties:
\begin{description}
\item[(C1)] (Symmetry) $\gsim(x,y)=\gsim(y,x)$ for all $x, y \in \Omega$.
\item[(C2)] (Null condition) For all $x \in \Omega$, $\sum_{y \in \Omega}\gsim (x,y)=0$.
\item[(C3)] (Nonnegativity) For all $x, y$ in $\Omega$,
\beq{cmeas1111}
\gsim(x,x)+\gsim (y,y) \ge 2\gsim (x, y).
\eeq
\end{description}

Moreover, we have
\beq{cind1111}
d(x,y)=(\gsim(x,x)+\gsim(y,y))/2 -\gsim(x,y).
\eeq
\iffalse
 an induced semi-metric and it can be viewed as the dual of the semi-cohesion measure.
On the other hand, if
$d(\cdot,\cdot)$ is a semi-metric for $\Omega$,
then
\bear{csim7777}
\gsim(x,y)&=&{1 \over n}\sum_{z_2 \in \Omega} d(z_2,y)+{1 \over n} \sum_{z_1 \in \Omega}d(x,z_1)\nonumber\\
&-&{1 \over n^2} \sum_{z_2 \in \Omega}\sum_{z_1 \in \Omega}d(z_2,z_1)-d(x,y).
\eear
is an induced semi-cohesion measure for $\Omega$.
\fi
Analogous to the argument in \cite{chang2015mathematical}, one can easily show the following
duality theorem.

\bthe{duality}
Consider a set of data points $\Omega$.
For a  semi-metric $d(\cdot,\cdot)$ that satisfies (D1)--(D3), let
\bear{dual1111}
d^*(x,y)&=&{1 \over n}\sum_{z_2 \in \Omega} d(z_2,y)+{1 \over n} \sum_{z_1 \in \Omega}d(x,z_1)\nonumber\\
&-&{1 \over n^2} \sum_{z_2 \in \Omega}\sum_{z_1 \in \Omega}d(z_2,z_1)-d(x,y)
\eear
be the induced semi-cohesion measure of $d(\cdot,\cdot)$. On the other hand, for a semi-cohesion measure
$\gsim(\cdot,\cdot)$ that satisfies (C1)--(C3), let
\beq{dual2222}
\gsim^*(x,y)=(\gsim(x,x)+\gsim(y,y))/2 -\gsim(x,y).
\eeq
Then $\gsim^*(x,y)$ is a semi-metric that satisfies (D1)--(D3). Moreover,
$d^{**}(x,y)=d(x,y)$ and $\gsim^{**}(x,y)=\gsim(x,y)$ for all $x, y \in \Omega$.
\ethe

In view of the duality result, there is a one-to-one mapping between a semi-metric and a semi-cohesion measure. Thus, we will simply say data points are in a semi-metric space if there is either a semi-cohesion measure or a semi-metric
associated with these data points.

\bsubsec{Clusters in a semi-metric space}{clusters}

In this section, we define what a cluster is for a set of data points in a semi-metric space.

\bdefin{cluster} {\bf (Cluster)}
Consider a set of $n$ data points, $\Omega=\{x_1, x_2, \ldots, x_n\}$, with a semi-cohesion measure $\gsim(\cdot,\cdot)$. For two sets $S_1$ and $S_2$, define
\beq{sum1234}
\gsim(S_1,S_2)=\sum_{x \in S_1} \sum_{y \in S_2} \gsim(x,y).
\eeq
Two sets $S_1$ and $S_2$ are said to be {\em cohesive} (resp. {\em incohesive}) if $\gsim(S_1, S_2) \ge 0$ (resp. $\gsim(S_1, S_2) \le 0$).
A nonempty set $S$ of $\Omega$ is called a {\em cluster} (with respect to the semi-cohesion measure $\gsim(\cdot,\cdot)$) if
\beq{cluster1111}
\gsim(S, S) \ge 0.
\eeq
\edefin

Following the same argument in \cite{chang2015mathematical}, one can also show a theorem for various equivalent statements for what a cluster is in a semi-metric space.
%The proofs are omitted.

\bthe{clustereq}
Consider a set of $n$ data points, $\Omega=\{x_1, x_2, \ldots, x_n\}$, with a semi-cohesion measure $\gsim(\cdot,\cdot)$.
Let $d(x,y)=(\gsim(x,x)+\gsim(y,y))/2 -\gsim(x,y)$ be the induced semi-metric and
\beq{avg1111}
\dbar(S_1, S_2) = {1 \over {|S_1| \times {|S_2|}}} \sum_{x \in S_1}\sum_{y \in S_2} d(x,y).
\eeq
be the average ``distance'' between  two randomly selected points with one point in $S_1$ and the other point in $S_2$.
Consider a {\em nonempty} set
 $S$ that is not equal to $\Omega$. Let $S^c = \Omega \backslash S$ be the set of points that are not in $S$.
The following statements are equivalent.
\begin{description}
\item[(i)] The set $S$ is a cluster, i.e., $\gsim(S,S)\ge 0$.
\item[(ii)] The set $S^c$ is a cluster, i.e., $\gsim (S^c, S^c) \ge 0$.
\item[(iii)] The two sets $S$ and $S^c$ are incohesive, i.e., $\gsim(S, S^c) \le 0$.
\item[(iv)] The set  $S$ is more cohesive to itself than to $S^c$, i.e., $\gsim(S,S) \ge \gsim(S,S^c)$.
\item[(v)] $2\dbar (S, \Omega)-\dbar(\Omega, \Omega)-\dbar (S,S)\ge 0$.
\item[(vi)] $2\dbar(S, S^c)- \dbar(S,S)-\dbar(S^c,S^c) \ge 0$.
\end{description}
\ethe

The condition for a cluster in \rthe{clustereq}(vi) is of particular importance as it allows us to characterize a cluster by using the average distance measures on the set $S$ and its complement $S^c$. Such a condition will be used for proving our main result in \rthe{main}.

\bsubsec{The \ksetsplus algorithm}{ksetsplus}

Though the extensions of the duality result and the equivalent statements for clusters to semi-metric spaces are basically the same as
those in \cite{chang2015mathematical}, one problem arises when extending the \ksets algorithm  to a semi-metric space. The key problem is that
the {\em triangular distance ($\Delta$-distance)} defined in the \ksets algorithm (see \rdef{kd} below)  might not be nonnegative in a semi-metric space.

\bdefin{kd}{\bf ($\Delta$-distance \cite{chang2015mathematical})} For a symmetric bivariate function $\gsim(\cdot,\cdot)$ on a set of data points $\Omega=\{x_1, x_2, \ldots, x_n\}$, the {\em $\Delta$-distance} from a point $x$ to a set $S$, denoted by $\Delta(x, S)$, is defined as follows:
\beq{kmeans5566}
\Delta(x, S)=\gsim(x,x)-\frac{2}{|S|} \gsim(x,S)+ \frac{1}{|S|^2}\gsim(S,S),
\eeq
where $\gsim(S_1, S_2)$ is defined \req{sum1234}.
%\beq{nmo2222}
%\gsim(S_1, S_2)=\sum_{x \in S_1} \sum_{y \in S_2} \gsim(x,y).
%\eeq
\edefin

Note from  \req{csim7777} that the $\Delta$-distance from a point $x$ to a set $S$ in a semi-metric space
can also be written as follows:
\beq{triang11110s}
\Delta(x, S)= {1 \over {|S|^2}}\sum_{z_1 \in S} \sum_{z_2 \in S}\Big (d(x,z_1)+d(x,z_2)-d(z_1, z_2)\Big ).
\eeq
Now consider the data set of three points $\Omega=\{x,y,z\}$ with the semi-metric $d(\cdot,\cdot)$ in Table \ref{tab:example}.
For $S=\{y,z\}$, one can easily compute from \req{triang11110s} that $\Delta(x,S)=-1<0$.

\begin{table}[ht]
\centering
\caption{A data set of three points $\Omega=\{x,y,z\}$ with a semi-metric $d(\cdot,\cdot)$.\label{tab:example}}
\begin{tabular}{|c|c|c|c|}
\hline
 $d(\cdot,\cdot)$                   & $x$ & $y$    & $z$  \\ \hline
$x$            & 0           & 1 & 1 \\ \hline
$y$            & 1           & 0 & 6 \\ \hline
$z$            & 1           & 6 & 0 \\ \hline
\end{tabular}
\end{table}

Since the $\Delta$-distance might not be nonnegative in a semi-metric space, the proofs for the convergence and the performance guarantee of the \ksets algorithm in \cite{chang2015mathematical} are no longer valid.
Fortunately, the $\Delta$-distance in a semi-metric space has the following (weaker) nonnegative property that will enable us to prove
 the performance guarantee of the  \ksetsplus algorithm (defined in Algorithm \ref{alg:ksetsplus} later) for clustering data points in a semi-metric space.

\bprop{nonnegative}
Consider a data set $\Omega=\{x_1, x_2, \ldots, x_n\}$ with a semi-metric $d(\cdot,\cdot)$. For any  subset $S$ of $\Omega$,
\beq{nonneg1111}
\sum_{x \in S}\Delta(x, S) =|S|\dbar(S,S) \ge 0.
\eeq
\eprop

The proof of \req{nonneg1111} in \rprop{nonnegative} follows directly from \req{triang11110s} and \req{avg1111}.
To introduce the \ksetsplus algorithm, we first define the adjusted $\Delta$-distance in \rdef{kdn} below.

\bdefin{kdn}{\bf (Adjusted  $\Delta$-distance)}
The {\em adjusted  $\Delta$-distance}   from a point $x$ to a set $S$, denoted by $\Deltaw(x, S)$, is defined as follows:
\beq{kd2222}
\Deltaw(x,S)=\left \{
\begin{array}{ll}
{{|S|} \over {|S|+1}} \Delta(x,S), & \mbox{if}\;x \not \in S, \\
{{|S|} \over {|S|-1}} \Delta(x,S), & \mbox{if}\;x \in S\;\mbox{and}\; |S|>1,\\
-\infty,& \mbox{if}\; x \in S\;\mbox{and}\;|S|=1.
                \end{array} \right.
\eeq
\edefin

Instead of using the  $\Delta$-distance for the assignment of a data point in the \ksets algorithm, we use the adjusted  $\Delta$-distance
for the assignment in the  \ksetsplus algorithm. We outline the \ksetsplus algorithm
 in Algorithm \ref{alg:ksetsplus}. Note that in Algorithm \ref{alg:ksetsplus}, the bivariate function $\gsim(\cdot,\cdot)$ is required to be symmetric, i.e., $\gsim(x,y)=\gsim(y,x)$. If $\gsim(\cdot,\cdot)$ is not symmetric, one may consider using $\hat \gsim(x,y)=(\gsim(x,y)+\gsim(y,x))/2$.

\begin{algorithm}[t]
%\SetAlgoNoLine
\KwIn{A data set $\Omega=\{x_1, x_2, \ldots, x_n\}$, a {\em symmetric} matrix $G=(\gsim(\cdot, \cdot))$ and the number of sets $K$.
}
\KwOut{A partition of sets $\{S_1, S_2, \ldots, S_K\}$.}

%\item[(P1)]
\noindent {\bf (0)} Initially, choose arbitrarily $K$ disjoint nonempty sets $S_1 , \ldots, S_K$ as a partition of $\Omega$.
%Let $c(x_i)$, $i=1,2, \ldots, n$, be the index of the set to which $x_i$ belongs, i.e., $x_i \in S_{c(x_i)}$.

%\noindent {\bf (1)} \While {{\em change==1}}{
%Set the flag {\em change $\leftarrow$ 0}\;
\noindent {\bf (1)} \For{$i=1, 2, \ldots, n$}{
%For $i=1, 2, \ldots, n$, do the following:

\noindent
Compute the adjusted $\Delta$-distance $\Deltaw(x_i, S_k)$ for each set $S_k$ by using \req{kmeans5566} and \req{kd2222}.
Find the set to which the point $x_i$ is closest in terms of the adjusted $\Delta$-distance. Assign that point $x_i$ to that set.}

\noindent {\bf (2)} Repeat from (1) until there is no further change.
\caption{The \ksetsplus Algorithm}
\label{alg:ksetsplus}
\end{algorithm}

In the following theorem, we show the convergence and the performance guarantee of the \ksetsplus algorithm.
The proof of \rthe{main} is given in Appendix A.

\bthe{main}
%{\bf (Pairwise interchange lemma)}
For a data set $\Omega=\{x_1, x_2, \ldots, x_n\}$ with a {\em symmetric}  matrix $G=(\gsim(\cdot, \cdot))$,
consider the clustering problem that finds  a partition $\{S_1, S_2, \ldots, S_K\}$ of $\Omega$ with a fixed $K$ that maximizes
the objective function
$\sum_{k=1}^K {1 \over {|S_k|}}\gsim(S_k,S_k)$.
\begin{description}
\item[(i)]
The \ksetsplus algorithm in Algorithm \ref{alg:ksetsplus} converges monotonically to a local optimum of the optimization problem in a finite number of iterations.
\item[(ii)] Suppose that $\gsim(\cdot, \cdot)$ is a semi-cohesion measure. Let $S_1, S_2, \ldots, S_K$ be the $K$ sets when the algorithm converges. Then for all $i \ne j$,  the   two sets $S_i$ and $S_j$ are two clusters if these two sets are viewed in isolation (by removing the data points not in $S_i \cup S_j$ from $\Omega$).
\end{description}
\ethe

In particular, if $K=2$, it then follows from \rthe{main}(ii) that the \ksetsplus algorithm yields two clusters for data points in a semi-metric space.

\bsec{Beyond semi-metric spaces}{beyond}

\bsubsec{Clustering with a symmetric similarity measure}{general}

In this section, we further extend the applicability of the \ksetsplus algorithm to the clustering problem with a symmetric similarity measure. A similarity measure is generally referred to as a bivariate function that measures how similar two data points are. The clustering problem with a similarity measure is to cluster data points so that similar data points are clustered together. For a symmetric similarity measure $\gsim(\cdot,\cdot)$,
we have shown in \rthe{main}(i) that the \ksetsplus algorithm in Algorithm \ref{alg:ksetsplus} converges monotonically to a local optimum of the optimization problem $\sum_{k=1}^K {1 \over {|S_k|}}\gsim(S_k,S_k)$ within a finite number of iterations. Thus, the \ksetsplus algorithm can be applied for clustering with a symmetric similarity measure. But what is the physical meaning of the sets returned by the \ksetsplus algorithm for such a symmetric similarity measure?
In order to answer this question, we show there is  a natural semi-cohesion measure from a symmetric similarity measure and
the \ksetsplus algorithm that uses this  semi-cohesion measure converges to the same partition as that using the original symmetric similarity measure (if they both use the same initial partition).
As a direct consequence of  \rthe{main}(ii), any two sets returned by the \ksetsplus algorithm for such a symmetric similarity measure are clusters with respect to the semi-cohesion measure  when they are viewed in isolation.

 In \rlem{precoh} below, we first show how one can map  a symmetric similarity measure to  a semi-cohesion measure.
The proof is given in Appendix B.
%can be done by  simple algebraic derivations  and thus omitted for space limitation.

\blem{precoh}
For a symmetric similarity  measure $\gsim(\cdot,\cdot)$,
let
\bear{precoh1111}
\tgsim(x,y)&=&\gsim(x,y)-{1 \over n}\gsim(x, \Omega)
-{1 \over n}\gsim(y, \Omega)\nonumber\\
& +&{1 \over {n^2}}\gsim(\Omega, \Omega)+ \sigma \delta(x,y)- {\sigma \over n},
\eear
where $\delta(x,y)$ is the usual $\delta$ function (that has value 1 if $x=y$ and 0 otherwise), and $\sigma$ is a constant that satisfies
\beq{precoh1122}
\sigma \ge \max_{x \ne y}[\gsim(x,y)-(\gsim(x,x)+\gsim(y,y))/2].
\eeq
Then
the bivariate function $\tgsim(\cdot,\cdot)$ in \req{precoh1122} is a {\em semi-cohesion measure} for $\Omega$, i.e.,
it satisfies (C1), (C2) and (C3).
\elem

In the following lemma, we further establish the connections for the $\Delta$-distance and the adjusted $\Delta$-distance
between the original symmetric similarity measure $\gsim(\cdot,\cdot)$ and the  semi-cohesion measure $\tgsim(\cdot,\cdot)$ in \req{precoh1111}.
The proof is given in Appendix C.
%can be done by simple algebraic derivations  and thus omitted for space limitation.

\blem{Shift}
Let $\Delta(x,S)$ (resp. ${\tilde \Delta}(x, S)$) be the $\Delta$-distance from a point $x$ to a set $S$ with respect to $\gsim(\cdot,\cdot)$ (resp. $\tgsim(\cdot,\cdot)$).
Also, let $\Deltaw(x, S)$ (resp. $\tDeltaw(x, S)$) be the adjusted $\Delta$-distance  from a point $x$ to a set $S$ with respect to $\gsim(\cdot,\cdot)$ (resp. $\tgsim(\cdot,\cdot)$).
Then
\beq{shift1111}
{\tilde \Delta}(x, S)=\left \{
\begin{array}{ll}
 \Delta(x,S)+\sigma (1- {1 \over {|S|}}), & \mbox{if}\;x \not \in S, \\
 \Delta(x,S)+\sigma (1+ {1 \over {|S|}}), & \mbox{if}\;x \in S.\\
                \end{array} \right.
\eeq
Moreover,
\beq{shift2222}
\tDeltaw(x, S)=\Deltaw(x, S)+\sigma.
\eeq
\elem

It is easy to see that for any partition $S_1, S_2, \ldots,, S_K$
\bear{twosetsgen1111}
&&\sum_{k=1}^K {1 \over {|S_k|}}\tgsim(S_k,S_k)\nonumber\\
&&=\sum_{k=1}^K {1 \over {|S_k|}}\gsim(S_k,S_k)-{1 \over n}\gsim(\Omega,\Omega)+(K-1)\sigma.
\eear
Thus, optimizing $\sum_{k=1}^K {1 \over {|S_k|}}\gsim(S_k,S_k)$ with respect to the symmetric similarity measure $\gsim(\cdot,\cdot)$ is equivalent to optimizing $\sum_{k=1}^K {1 \over {|S_k|}}\tgsim(S_k,S_k)$ with respect to the semi-cohesion measure $\tgsim(\cdot,\cdot)$. Since
\beq{shift2222b}
\tDeltaw(x, S)=\Deltaw(x, S)+\sigma,
\eeq
we conclude that for these two optimization problems
the \ksetsplus algorithm converges to the same partition  if they both use the same initial partition.

Note that the K-means algorithm needs the data points to be in a Euclidean space, the kernel K-means algorithm
needs the data points to be mapped into some Euclidean space, and the \ksets algorithm
needs the data points to be in a metric space. The result in \rlem{Shift} shows that the  \ksetsplus algorithm
lifts all the constraints on the data points and it can be operated merely by a symmetric similarity measure.

\bsubsec{Computational complexity}{complexity}

In this section, we address  the computational complexity and the memory complexity of the \ksetsplus algorithm.
For an $n \times n$ symmetric similarity matrix with only $m$ nonzero elements in the matrix,
we show that
 the computational complexity of the  \ksetsplus algorithm  is  $O((Kn+m)I)$, where $I$ is the number of iterations. The memory complexity to achieve that computational complexity is $O(Kn+m)$.

Note that the main computation overhead of the \ksetsplus algorithm is mainly for the computation of the adjusted $\Delta$-distance.  In view of  \req{kmeans5566}, we know that one needs to compute $\gsim(x,S)$ and ${1 \over {|S|^2}}\gsim(S,S)$ in order to compute
$\Delta(x, S)$.
Let
\beq{sparse1111}
\gbar(S_1, S_2)={1 \over {|S_1| |S_2|}} \gsim(S_1,S_2).
\eeq
Our approach to reduce the computational complexity is to store $\gbar(S_k,S_k)$, $k=1,2, \ldots, K$ and $\gsim(x_i,S_k)$ for $i=1,2, \ldots, n$ and $k=1,2, \ldots, K$. Once these are stored in memory, one can compute the adjusted $\Delta$-distance $\Deltaw(x_i,S_k)$ in $O(1)$ steps.
Suppose that $x_i$ is originally in the set $S_{\seta}$ and it is reassigned to $S_{\setb}$.
Then $\gbar(S_{\setb} \cup\{x_i\},S_{\setb} \cup \{x_i\})$ can be updated by computing
\begin{eqnarray}&&{{|S_{\setb}|^2} \over {(|S_{\setb}|+1)^2}}\gbar(S_{\setb}, S_{\setb})+ {{2|S_{\setb}|} \over {(|S_{\setb}|+1)^2}}\gbar(\{x_i\}, S_{\setb})\nonumber \\&&\quad\quad\quad\quad+{1 \over {{(|S_{\setb}|+1)^2}}}\gsim(x_i,x_i).
\label{eq:temp1111}
\end{eqnarray}
Also, $\gbar(S_{\seta} \backslash \{x_i\},S_{\seta} \backslash \{x_i\})$ can be updated by computing
\begin{eqnarray}
&&{{|S_{\seta}|^2} \over {(|S_{\seta}|-1)^2}}\Big (\gbar(S_{\seta},S_{\seta})-2{{1} \over {|S_{\seta}|}}\gbar(\{x_i\},S_{\seta}) \Big )\nonumber \\&&\quad\quad\quad\quad+{1 \over{{(|S_{\seta}|-1)^2}}}\gsim(x_i,x_i).
\label{eq:temp2222}
\end{eqnarray}
Such updates can be done in $O(1)$ steps.
On the other hand, let
\beq{sparse2222}\mbox{Nei}(i)=\{j: \gsim(x_i,x_j) \ne 0\}
 \eeq
 be the set of data points that are neighbors of $x_i$. Note that if $\gsim(x_i, x_i) \ne 0$, then $x_i$ is also in  $\mbox{Nei}(i)$. When $x_i$ is moved from $S_{\seta}$ to $S_{\setb}$,
we only need to update $\gsim(y,S_{\seta})$ and $\gsim(y,S_{\setb})$ for the data point $y$ that is a neighbor of $x_i$. Specifically,
For each node $y \in \mbox{Nei}(i)$, update
\bearn
&&\gsim(y,S_{\setb})  \leftarrow  \gsim(y,S_{\setb}) +\gsim(y,x_i), \\
&&\gsim(y, S_{\seta})  \leftarrow  \gsim(y, S_{\seta}) -\gsim(y,x_i).
\eearn
Such updates can be done in $O(\mbox{Nei}(i))$ steps.
Let
\beq{edges1111}
m=\sum_{i=1}^n|\mbox{Nei}(i)|
\eeq
 be the total number of nonzero entries in the $n \times n$ symmetric matrix $G=(\gsim(\cdot, \cdot))$.
Then the total number of updates for all the data points $x_i$, $i=1,2, \ldots, n$, can be done in $O(m)$ steps.
%In Algorithm \ref{alg:sparse}, we outline the pseudo code for the  \ksetsplus algorithm of this implementation.
Since we need to compute the $\Delta$-distance for the $K$ sets for each data point in the {\em for} loop in  Algorithm \ref{alg:ksetsplus},
 the computational complexity of the  \ksetsplus algorithm of this implementation is thus $O((Kn+m)I)$, where $I$ is the number of iterations in the {\em for} loop of Algorithm \ref{alg:ksetsplus}. Regarding the memory complexity, one can store the symmetric matrix $G=(\gsim(\cdot, \cdot))$ in the adjacency list form and that requires $O(m)$ amount of memory. The memory requirement for storing $S_k$,  $k=1,2, \ldots, K$, $\gbar(S_k,S_k)$, $k=1,2, \ldots, K$ and $\gsim(x_i,S_k)$ for $i=1,2, \ldots, n$ and $k=1,2, \ldots, K$ is $O(Kn)$. Thus, the overall memory complexity is $O(Kn+m)$.

\bsec{Experiments}{exp}

In this section, we
evaluate the performance of the \ksetsplus algorithm by conducting two experiments: (i) community detection of signed networks generated by the stochastic block model in Section \ref{sec:signed}, and (ii)
 clustering of a real network   from the WonderNetwork website \cite{Wonder} in Section \ref{sec:router}.

\bsubsec{Community detection of signed networks with two communities}{signed}

In this section, we conduct experiments for the \ksetsplus algorithm by using the signed networks from the stochastic block model. We follow the procedure in \cite{INFOCOM2017} to generate the test networks.
Each test network consists of
$n$ nodes and two ground-truth blocks, each with $n/2$ nodes.
There are three key parameters $p_{in}$, $p_{out}$, and $p$ for generating a test network.
The parameter $p_{in}$ is the probability that there is a {\em positive} edge between two nodes within the same block and $p_{out}$ is the probability that there is a {\em negative} edge between two nodes in two different blocks. All edges are generated independently according to $p_{in}$ and $p_{out}$.  After all the signed edges are generated,
we then  flip the sign of an edge independently with the crossover probability $p$.

In our experiments, the total number of nodes in the stochastic block model is $n=2000$ with 1000 nodes in each block. Let $c=(n/2-1)p_{in}+np_{out}/2$ be the average degree of a node, and it is set to be 6, 8, and 10, respectively. Also, let $c_{in}=np_{in}$ and $c_{out}=np_{out}$. The value of $c_{in}-c_{out}$ is set to be 5 and that is used with the average degree $c$ to uniquely determine $p_{in}$ and $p_{out}$.
The crossover probability $p$ is in the range from 0.01 to 0.2 with a common step of 0.01. We generate 20 graphs for each $p$ and $c$. We remove isolated nodes, and thus the exact numbers of nodes in the experiments might be less than 2000. We show the experimental results with each point averaged over 20 random graphs. The error bars represent the 95\% confident intervals.

 To test the \ksetsplus algorithm, we use the similarity matrix $G$ with
 \beq{BF11b}
G=A+0.5A^2,
\eeq
where $A$ is the adjacency matrix of the signed network after randomly flipping the sign of an edge.
Such a similarity matrix was suggested in \cite{INFOCOM2017} for community detection in signed networks as
it allows us to ``see'' more than one step relationship between two nodes.

\begin{figure}[h]
\centering
\includegraphics[width=0.45\textwidth]{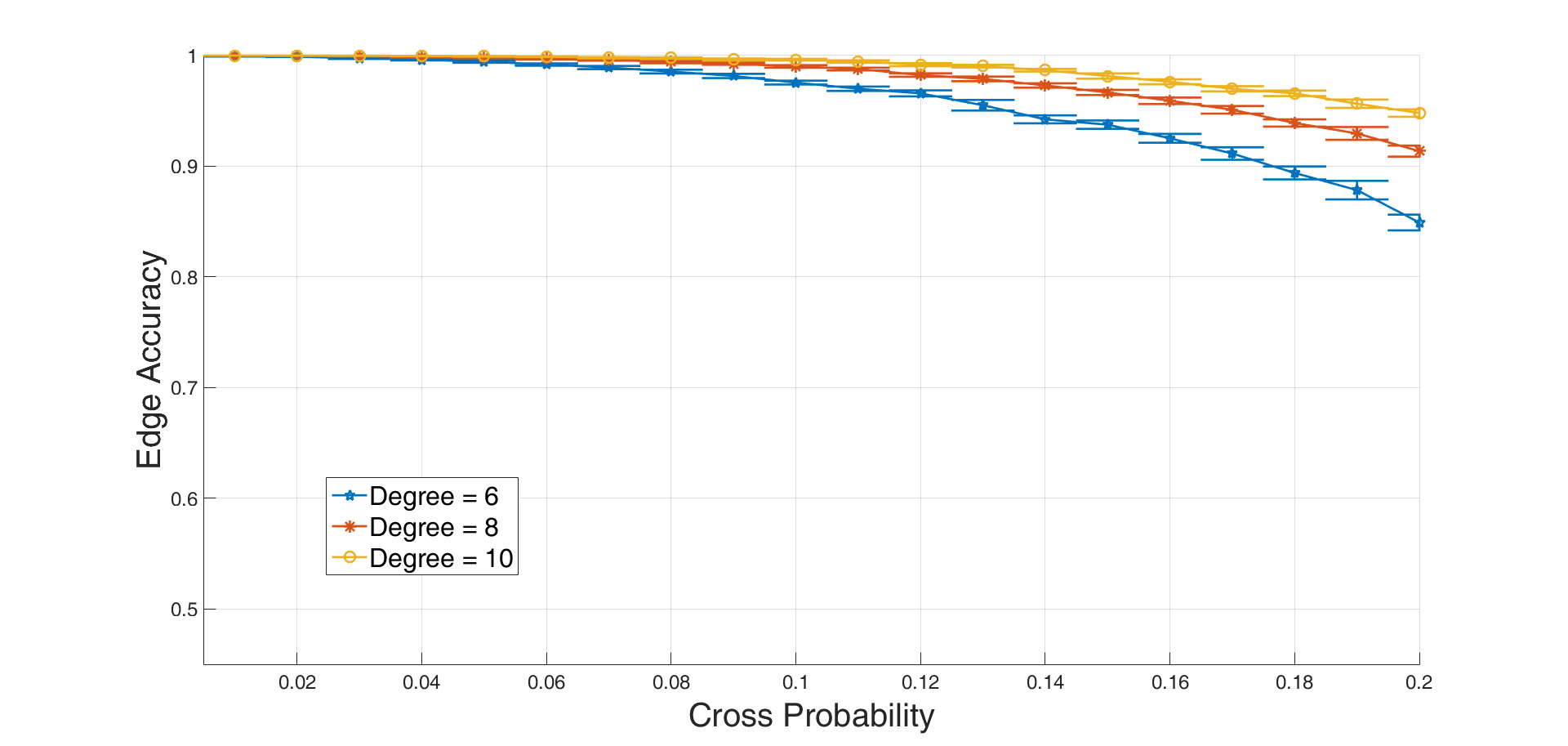}
\caption{Community detection of signed networks with two communities.}
\label{fig:accuracy}
\end{figure}

In \rfig{accuracy}, we show our experimental results for edge accuracy (the percentage of edges that are correctly detected) as a function of the crossover probability $p$. As shown in \rfig{accuracy}, the \ksetsplus algorithm performs very well. For $c=10$, it can still recover almost 100\% of the edges even when the crossover probability $p$ is 0.1 and roughly 95\% of the edges when the crossover probability $p$ is 0.2.
Also, increasing the average degree $c$ in
the stochastic block model also increases the edge accuracy for
the \ksetsplus algorithm.
This might be due to the fact that the tested signed networks with a larger average degree
are more dense.

\bsubsec{Clustering of a real network}{router}

\begin{figure}[tb]
    \begin{center}
%    \begin{tabular}{p{0.3\textwidth}p}
      \includegraphics[width=0.5\textwidth]{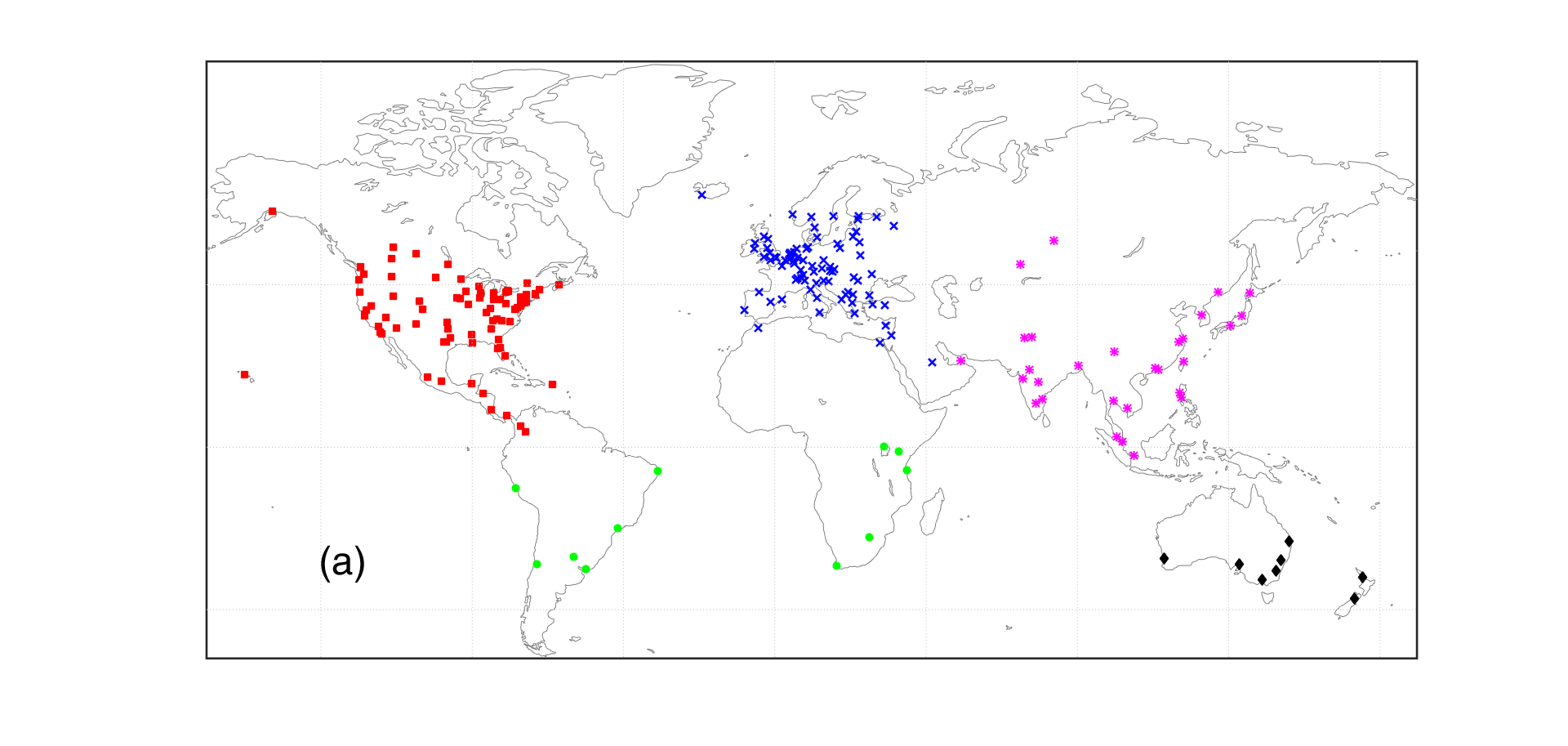} \\
      \includegraphics[width=0.5\textwidth]{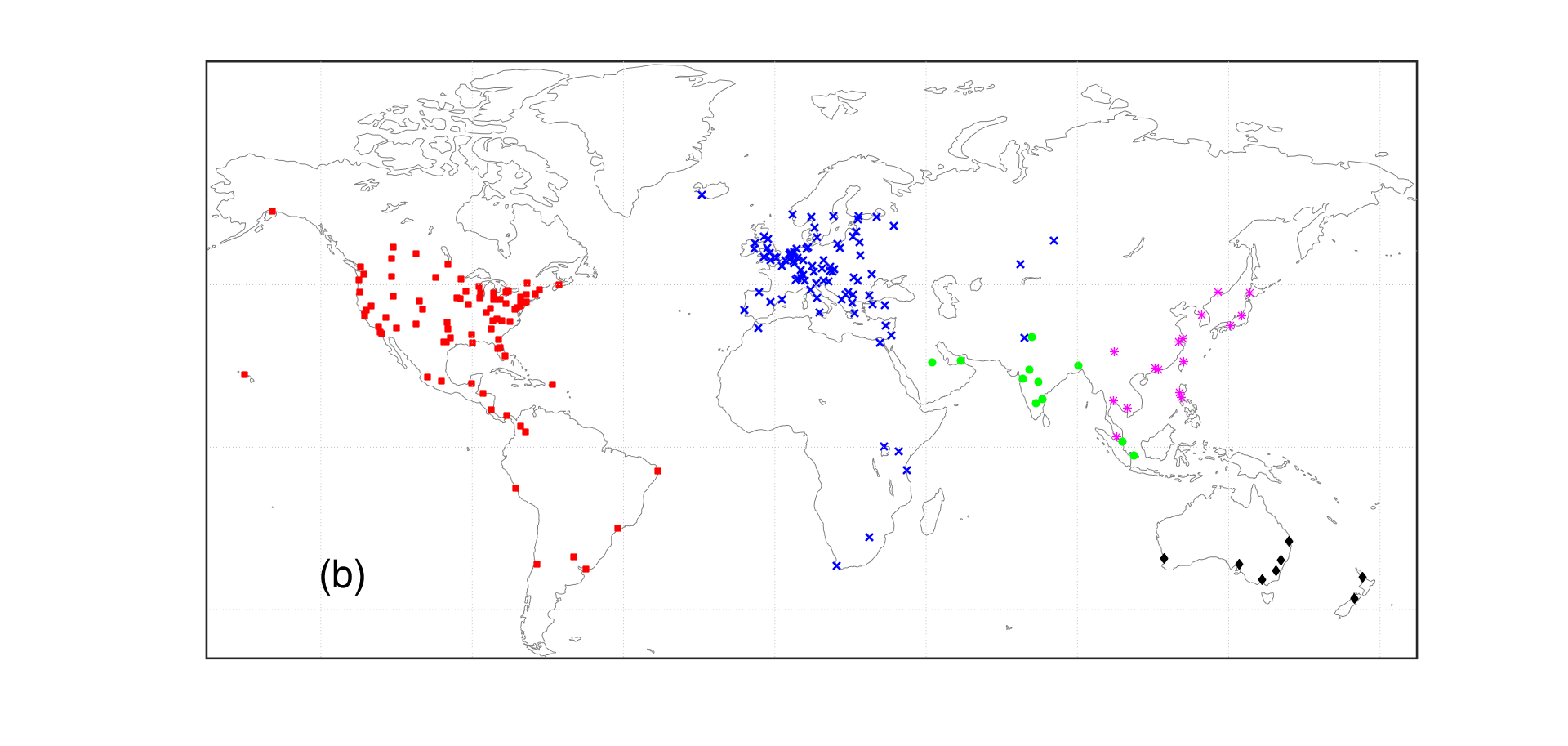}\\
%       \\ \\
%      \end{tabular}
    \caption{Clustering for the WonderNetwork dataset: (a) the (geographic) distance matrix and (b) the latency matrix.}
    \label{fig:Wonder}
  \end{center}
\end{figure}

In this section, we test the \ksetsplus algorithm on the real  network  from the WonderNetwork website \cite{Wonder}. In this dataset,  there are 216 servers in different locations and the latency (measured by the round trip time) between any two servers of these 216 servers are recorded in real time. The dataset in our experiment is a snapshot on Sept. 24, 2016. For this dataset, the triangular inequality is not always satisfied. For example, we notice that latency(Adelaide, Athens)=250, latency(Athens, Albany)=138, latency(Adelaide, Albany)=400, and $250+138 \le 400$. In addition to  the latency data, the WonderNetwork website also provides the geographic location of each server. We then use the Haversine formula to compute the  distance  between any two servers.
In the WonderNetwork dataset, the latency measure from location $L_1$ to location $L_2$ is  slightly different from that from location $L_2$ to location $L_1$. To ensure that the latency measure is a semi-metric, we simply symmetrize the latency matrix by taking the average of the latency measures from both directions. In our experiments, the number of clusters $K$ is set to 5. We run 20 times of the \ksetsplus algorithm by using the distance matrix and the latency matrix, respectively.
In each of the 20 trials, the initial partition  is randomly selected. The output partition that has the best objective value from these 20 trials is selected. The results for the distance matrix and the latency matrix are shown in \rfig{Wonder}(a) and (b), respectively.
In \rfig{Wonder}(a) and (b), the servers that are in the same cluster are marked with the same colored marker.
In view of \rfig{Wonder}(a), we can observe that almost all the servers are partitioned into densely packed clusters except for the servers in South America and Africa. On the other hand, as shown in \rfig{Wonder}(b), the servers in South America and Africa are  merged into other clusters. To shed more light on these interesting differences, we  compare the findings in \rfig{Wonder}(b) to the Submarine Cable Map \cite{Submarine} (which records the currently active submarine cables). We notice that there are many cables placed around South America, connecting to the Caribbean and then to the East Coast of the United States. These cables greatly reduce the latency from South America to North America and thus cause the servers in South America to be clustered with the servers in North America. Similarly, there are a few connected cables between Africa and Europe. Therefore, the servers in Africa and Europe are clustered together. Due to many directly connected cables  from Dubai to the Singapore Strait, servers around India are not clustered with the other servers in Asia. In particular, there are two servers marked with green dots, located in Jakarta and Singapore, that are clustered with servers in India even though they are geographically  closer to the East Asia. These “outliers” have low latency to communicate with those servers in India. Finally, there are three servers, two in Russia and one in Lahore, that have low latency to the servers in Europe and they are clustered with the servers in Europe.
\iffalse
The result from our experiment shows the capability of K-sets+ algorithm for clustering data in some semi-metric space, which may be relevant to real-world problems. For example, we can use the result from figure (b) to provide a peer-to-peer content delivery network service. Since the K-sets+ algorithm minimizes the delta distance from each data point to the cluster it belongs to, the servers in the same cluster would have low communication latency between each other, helping each peer to query the content it need.
\fi

\bsec{Conclusion}{conclusion}

In this paper, we proposed  the \ksetsplus algorithm for clustering data points in a semi-metric space and data points that only have a symmetric similarity measure.  We showed that the \ksetsplus algorithm converges in a finite number of iterations and it retains the same performance guarantee as the \ksets algorithm in \cite{chang2015mathematical}. 
Moreover,  both the computational complexity and the memory complexity are linear in $n$ when the $n \times n$ similarity matrix is {\em sparse}, i.e., $m=O(n)$.
To show the effectiveness of the \ksetsplus algorithm,  we also conducted various experiments  by using a synthetic dataset from the stochastic block model and a real network from the WonderNetwork website \cite{Wonder}.

%\newpage

%\appendix
\section*{Appendix A}

\setcounter{section}{1}

In this section, we prove \rthe{main}.

\noindent (i)
It suffices to show that if $x$ is in a set $S_1$ with $|S_1|>1$ and $\Deltaw(x, S_2) < \Deltaw(x, S_1)$, then move point $x$ from $S_1$ to $S_2$ increases the value of the objective function.
Let $S_k$ (resp. $S^\prime_k$), $k=1,2, \ldots, K$, be the partition before (resp. after) the change.
Also let $R$ (resp. $R^\prime$) be the value of the objective function  before (resp. after) the change.
Then
\bearn
&&R^\prime -R\nonumber\\
 &&={{\gsim(S_1 \backslash \{x\},S_1 \backslash \{x\})} \over {|S_1|-1}}+{{\gsim(S_2 \cup \{x\},S_2 \cup \{x\})} \over {|S_2|+1}}\nonumber\\
 &&\quad
- {{\gsim(S_1 ,S_1 )} \over {|S_1|}}-{{\gsim(S_2 ,S_2) } \over {|S_2|}}
.
\eearn
Since
\bearn
&&\gsim(S_2 \cup \{x\},S_2 \cup \{x\})\\
&&=\gsim(S_2, S_2)+2 \gsim(x, S_2)+\gsim(x,x),
\eearn
we have from \req{kmeans5566} and \req{kd2222} that
\bearn
&&{{\gsim(S_2 \cup \{x\},S_2 \cup \{x\})} \over {|S_2|+1}}-{{\gsim(S_2 ,S_2) } \over {|S_2|}}\\
&&={{2|S_2|\gsim(x, S_2)+|S_2|\gsim(x, x)-\gsim(S_2,S_2)} \over {|S_2| \cdot (|S_2|+1)}} \\
&&=\gsim(x,x)-{{|S_2|} \over {|S_2|+1}} \Delta (x,S_2)\\
&&=\gsim(x,x)-\Deltaw(x,S_2).
\eearn
On the other hand,
we note that
\beq{kd1234}
\gsim(S_1 \backslash \{x\},S_1 \backslash \{x\})=\gsim(S_1,S_1)-2\gsim(x, S_1)+\gsim(x,x).
\eeq
Using \req{kd1234}, \req{kmeans5566} and \req{kd2222} yields
\bearn
&&{{\gsim(S_1 \backslash \{x\},S_1 \backslash \{x\})} \over {|S_1|-1}}-{{\gsim(S_1 ,S_1 )} \over {|S_1|}}\\
&&={{-2|S_1|\gsim(x, S_1)+|S_1|\gsim(x,x)+\gsim(S_1, S_1)} \over {|S_1| \cdot (|S_1|-1)}}\\
&&=-\gsim(x,x)+{{|S_1|} \over {|S_1|-1}} \Delta (x,S_1)\\
&&=-\gsim(x,x)+\Deltaw(x,S_1).
\eearn
Thus,
$$R^\prime -R =\Deltaw(x,S_1)-\Deltaw(x,S_2)>0.$$

As the objective value is non-increasing after a change of the partition, there is no loop in the algorithm.
Since the number of partitions is finite, the algorithm thus converges in a finite number of steps (iterations).

\noindent (ii)
Let $d(\cdot,\cdot)$ be the induced semi-metric.
In view of \rthe{clustereq}(vi), it suffices to show that for all $i \ne j$
\beq{mainb1111}
2\dbar(S_i, S_j)- \dbar(S_i,S_i)-\dbar(S_j,S_j) \ge 0.
\eeq
If the set $S_i$ contains a single element $x$,  then
$$\dbar (S_i, S_i)=d(x,x)=0.$$
Thus,
the inequality in \req{mainb1111} holds trivially if $|S_i|=|S_j|=1$.

Now suppose that $\min(|S_i|, |S_j|) \ge 2$. Without loss of generality,
we assume that $|S_i|\ge 2$.
When the \ksetsplus algorithm converges, we know that for any $x \in S_i$,
$$\Deltaw(x, S_i) \le \Deltaw (x, S_j).$$
 Summing over $x \in S_i$ yields
\beq{two1111a}
\sum_{x \in S_i}\Deltaw(x, S_i) \le \sum_{x \in S_i}\Deltaw(x, S_j).
\eeq
Note from \req{kd2222} that for any $x \in S_i$,
$$\Deltaw(x, S_i)={{|S_i|} \over {|S_i|-1}} \Delta(x, S_i),$$
and
$$\Deltaw(x, S_j) ={{|S_j|} \over {|S_j|+1}}\Delta(x, S_j).$$
Thus, it follows from \req{two1111a} that
\beq{two1122a}
{{|S_i|} \over {|S_i|-1}}\sum_{x \in S_i}\Delta(x, S_i) \le {{|S_j|} \over {|S_j|+1}}\sum_{x \in S_i}\Delta(x, S_j).
\eeq
Note from \req{triang11110s} that
\beq{two2222a}\sum_{x \in S_i}\Delta(x, S_i)=|S_i| \dbar(S_i, S_i),
\eeq
and that
\beq{two3333a}
\sum_{x \in S_i}\Delta(x, S_j)=|S_i|(2 \dbar(S_i, S_j)-\dbar(S_j,S_j)).
\eeq
Since $d(\cdot,\cdot)$ is the induced semi-metric, we know from \rprop{nonnegative} that
$$\sum_{x \in S_i}\Delta(x, S_i) \ge 0.$$
Using this in \req{two1122a} yields
$$\sum_{x \in S_i}\Delta(x, S_j) \ge 0.$$
Thus, we have from \req{two1122a} that
\bear{two3366a}
&&\sum_{x \in S_i}\Delta(x, S_i) \le {{|S_i|} \over {|S_i|-1}}\sum_{x \in S_i}\Delta(x, S_i) \nonumber\\
&&\le {{|S_j|} \over {|S_j|+1}}\sum_{x \in S_i}\Delta(x, S_j) \le \sum_{x \in S_i}\Delta(x, S_j).
\eear
That the inequality in \req{mainb1111} holds follows directly from \req{two2222a}, \req{two3333a} and \req{two3366a}.

%\eproof

%\iffalse
\section*{Appendix B}

In this section, we prove \rlem{precoh}.

%\bproof
Since $\gsim(\cdot,\cdot)$ is symmetric, clearly $\tgsim(\cdot,\cdot)$ is also symmetric. Thus, (C1) is satisfied trivially.
To see that (C2) is satisfied, observe from \req{precoh1111} that
\bear{precoh2222}
\sum_{y \in \Omega}\tgsim(x,y)&=&\gsim(x,\Omega)-\gsim(x, \Omega)
-{1 \over n}\gsim(\Omega, \Omega) \nonumber\\ &+&{1 \over {n}}\gsim(\Omega, \Omega)+\sigma-\sigma =0.
\eear
To see that (C3) holds, we note that
\bear{precoh2255}
\tgsim(x,x)=\gsim(x,x)-{2 \over n}\gsim(x, \Omega)  +{1 \over {n^2}}\gsim(\Omega, \Omega)+{{(n-1)} \over n}\sigma,
\eear
and that
\bearn
\tgsim(y,y)=\gsim(y,y)-{2 \over n}\gsim(y, \Omega)  +{1 \over {n^2}}\gsim(\Omega, \Omega)+{{(n-1)} \over n}\sigma.
\eearn
Thus, for $x \ne y$, we have from \req{precoh1122} that
\bearn
&&\tgsim(x,x)+\tgsim(y,y)-2\tgsim(x,y) \nonumber \\
&&=\gsim(x,x)+\gsim(y,y)-2\gsim(x,y)+2 \sigma \ge 0.
\eearn
%\eproof

\section*{Appendix C}

In this section, we prove \rlem{Shift}.

%\bproof
Note from \rdef{kd} and \rdef{kdn} that
\beq{kmeans5566t}
{\tilde \Delta}(x, S)=\tgsim(x,x)-{2 \over |S|} \tgsim(x,S)+ {1 \over {|S|^2}}\tgsim(S,S),
\eeq
and that
\beq{kd2222s}
\tDeltaw(x,S)=\left \{
\begin{array}{ll}
{{|S|} \over {|S|+1}} {\tilde \Delta}(x,S), & \mbox{if}\;x \not \in S, \\
{{|S|} \over {|S|-1}} {\tilde \Delta}(x,S), & \mbox{if}\;x \in S\;\mbox{and}\; |S|>1,\\
-\infty,& \mbox{if}\; x \in S\;\mbox{and}\;|S|=1.
                \end{array} \right.
\eeq
To show \req{shift1111}, we need to consider two cases: (i) $x \in S$ and (ii) $x \not \in S$.
For both cases, we have from \req{precoh1111} that
\bear{precoh2255s}
\tgsim(x,x)=\gsim(x,x)-{2 \over n}\gsim(x, \Omega)  +{1 \over {n^2}}\gsim(\Omega, \Omega)+{{(n-1)} \over n}\sigma,
\eear
and
\bear{precoh4444s}
\tgsim(S,S)&=&\gsim(S,S)-{{|S|} \over n}\gsim(S, \Omega)
-{{|S|} \over n}\gsim(S, \Omega)\nonumber \\&+& {{|S|^2} \over {n^2}}\gsim(\Omega, \Omega)+\sigma |S| -\sigma {{|S|^2} \over n}.
\eear

Now we consider the first case that $x \in S$.
In this case,
note that for $x \in S$
\bear{precoh3333s}
\tgsim(x,S)&=&\gsim(x,S)-{{|S|} \over n}\gsim(x, \Omega)
-{{1} \over n}\gsim(S, \Omega) \nonumber \\&+& {{|S|} \over {n^2}}\gsim(\Omega, \Omega)+\sigma-\sigma{{|S|} \over n}.
\eear

Using \req{precoh2255s}, \req{precoh3333s} and \req{precoh4444s}  in \req{kmeans5566t}
yields
\bear{kmeans5566tt}
{\tilde \Delta}(x, S)&=&\tgsim(x,x)-{2 \over |S|} \tgsim(x,S)+ {1 \over {|S|^2}}\tgsim(S,S) \nonumber\\
&=&\gsim(x,x)-{2 \over |S|} \gsim(x,S)+ {1 \over {|S|^2}}\gsim(S,S)\nonumber \\
&\quad&+\sigma (1- {1 \over {|S|}})\nonumber \\
& =&\Delta (x,S)+\sigma (1- {1 \over {|S|}}).
\eear
From \req{kd2222s}, it then follows that
\bear{preco6666}
&&\tDeltaw(x,S)={{|S|} \over {|S|-1}}{\tilde \Delta}(x,S) \nonumber \\
&&={{|S|} \over {|S|-1}}\Big (\Delta (x,S)+\sigma (1- {1 \over {|S|}})\Big) \nonumber \\
&&=\Deltaw(x,S)+\sigma.
\eear

Now we consider the second case that $x \not \in S$.
In this case,
note that for $x \not \in S$
\bear{precoh3333b}
\tgsim(x,S)&=&\gsim(x,S)-{{|S|} \over n}\gsim(x, \Omega)
-{{1} \over n}\gsim(S, \Omega) \nonumber \\&+& {{|S|} \over {n^2}}\gsim(\Omega, \Omega)-\sigma{{|S|} \over n}.
\eear
Using \req{precoh2255s}, \req{precoh3333b} and \req{precoh4444s}  in \req{kmeans5566t}
yields
\bear{kmeans5566ttb}
{\tilde \Delta}(x, S)&=&\tgsim(x,x)-{2 \over |S|} \tgsim(x,S)+ {1 \over {|S|^2}}\tgsim(S,S) \nonumber\\
&=&\gsim(x,x)-{2 \over |S|} \gsim(x,S)+ {1 \over {|S|^2}}\gsim(S,S)\nonumber \\
&\quad&+\sigma (1+ {1 \over {|S|}})\nonumber \\
& =&\Delta (x,S)+\sigma (1+ {1 \over {|S|}}).
\eear
From \req{kd2222s}, it then follows that
\bear{preco6666b}
&&\tDeltaw(x,S)={{|S|} \over {|S|+1}}{\tilde \Delta}(x,S) \nonumber \\
&&={{|S|} \over {|S|+1}}\Big (\Delta (x,S)+\sigma (1+ {1 \over {|S|}})\Big) \nonumber \\
&&=\Deltaw(x,S)+\sigma.
\eear
%\eproof
%\fi

%\bibliographystyle{unsrt}
%\bibliographystyle{IEEEtran}
%\bibliography{bibliographyJACM}


\end{document}